\documentclass[twocolumn,superscriptaddress,longbibliography]{revtex4-1}
\usepackage{amssymb}
\usepackage{mathtools}
\usepackage[colorlinks=true,urlcolor=blue,citecolor=blue]{hyperref}
\usepackage{changes}
\usepackage{bbm, dsfont}
\usepackage{enumerate}

\newcommand{\R}{\mathbbm{R}}

\newcommand{\tr}{\operatorname{tr}}

\newcommand{\sign}{\text{sign}}

\newcommand{\interactiongraphG}{\mathcal{G}}
\newcommand{\interactiongraphVset}{\mathcal{V}}
\newcommand{\interactiongraphEset}{\mathcal{E}}
\newcommand{\depgraphG}{G}

\usepackage{changes}

\begin{document}
\title{Verifying the output of quantum optimizers with ground-state energy lower bounds}
\author{Flavio Baccari}
\affiliation{ICFO-Institut de Ciencies Fotoniques, The Barcelona Institute of Science and Technology, 08860 Castelldefels (Barcelona), Spain}
\affiliation{Max-Planck-Institut f\"ur Quantenoptik, Hans-Kopfermann-Stra{\ss}e 1, 85748 Garching, Germany}
\author{Christian Gogolin}
\affiliation{ICFO-Institut de Ciencies Fotoniques, The Barcelona Institute of Science and Technology, 08860 Castelldefels (Barcelona), Spain}
\author{Peter Wittek}
\thanks{In memoriam of Peter Wittek, who was a source of inspiration for all of us.}
\affiliation{Rotman School of Management, University of Toronto, M5S 3E6 Toronto, Canada}
\affiliation{Creative Destruction Lab, M5S 3E6 Toronto, Canada}
\affiliation{Vector Institute for Artificial Intelligence, M5G 1M1 Toronto, Canada}
\affiliation{Perimeter Institute for Theoretical Physics, N2L 2Y5 Waterloo, Canada}
\author{Antonio Ac\'in}
\affiliation{ICFO-Institut de Ciencies Fotoniques, The Barcelona Institute of Science and Technology, 08860 Castelldefels (Barcelona), Spain}
\affiliation{ICREA, Passeig Lluis Companys 23, 08010 Barcelona, Spain}

\begin{abstract}
  Solving optimisation problems encoded in the ground state of classical-spin systems is a focus area for quantum computing devices, providing upper bounds to the unknown solution. To certify these bounds, they are compared to those obtained by classical methods. However, even if the quantum bound beats them, this says little about how close it is to the unknown solution. We consider the use of relaxations to the ground-state problem as a benchmark for the output of quantum optimisers. These relaxations are radically more informative because they provide lower bounds to the ground-state energy.
The chordal branch and bound algorithm we present provides a series of systematically improving confidence regions where the ground-state energy provably lies. Interestingly, each step in the process requires only an effort polynomial in the system size.
Additionally, the algorithm exploits the locality and sparsity of relevant Ising spin models in a systematic way. This yields certified solutions for many of the problems that are currently addressed by heuristic optimisation algorithms more efficiently and for larger system sizes.
We apply the method to verify the output of a D-Wave 2000Q device and identify instances where the annealer does not reach the ground-state energy and, more importantly, instances where it does, something impossible to do by means of standard variational approaches. Our work provides a flexible and scalable method for the verification of the outputs produced by quantum optimization devices.
\end{abstract}
\maketitle

\section{Introduction}
Classical Ising models are among the most paradigmatic and widely studied models in statistical physics.
They are capable of describing an immense variety of interesting physics, ranging from ferromagnetic to frustrated and glassy phases.
Moreover, they are important in fields as diverse as risk assessment in finance, logistics, machine learning~\cite{Koller2007Graphical}, and image de-noising~\cite{bhatt1994robust} because the solution of many optimisation and decision problems, such as partitioning, covering, and satisfiability can be encoded in the ground state of such models~\cite{Lucas2014}.
Their generality and the exponentially growing spaces of spin configurations, however, preclude the existence of any efficient general purpose algorithm to obtain the ground state.
It is hence no surprise that a wealth of approximate but more scalable classical techniques for the energy minimisation in such models have been developed.

Recently, novel approaches that leverage the power of near-term quantum devices such as quantum annealers, variational quantum eigensolvers~\cite{Peruzzo2014}, variational circuits \cite{farhi2014qaoa,crooks2018performance} or networks of degenerate optical parametric oscillators~\cite{qnncloud2017qnn}, are proposed for performing such tasks~\cite{denchev2016what,Crosson2016,mandra2018deceptive}.
The quality of their outputs is usually benchmarked against some of the most scalable classical approaches, e.g., simulated annealing~\cite{Kirkpatrick1983} or variational ansatz classes based on tensor-network states~\cite{TNOrus}.
All of these methods share one common feature: they only provide upper bounds on the ground-state energy. 
On the one hand, this feature limits the verification power of these approaches, which are only able to identify instances where quantum devices do not reach the ground-state energy. 
On the other hand, even when a quantum optimiser beats all classical variational methods, there is no way to know if the output of the quantum device is actually close to the true ground-state energy, unless the test is performed on problems for which the solution is already known~\cite{hen2015probing}.
To overcome these limitations, it is important to develop schemes that provide reliable lower bounds to the ground-state energy of spin problems, against which the results of quantum devices, and also classical variational methods, can be compared.

In this work, we tackle this issue by leveraging relaxations of polynomial optimisation problems through semidefinite programming (SDP).
The proposed method provides lower bounds on the ground-state energy by optimizing over a larger set than the physical spin configurations.
We improve the scalability of this type of relaxations, by making use of a method known as the chordal extension, which allows us to exploit the physical locality and sparsity structure present in relevant problem instances.
All in all, this yields an increasingly precise hierarchy of rigorous lower and, in fact, also upper bounds on the ground-state energy.
Combining these bounds we obtain a scalable and flexible method that provides with polynomial effort a confidence region inside which the ground state energy provably lies. By making use of a branch-and-bound scheme, the confidence region can be systematically improved.  Although the complexity of the general Ising problem implies that one might have to run an exponential number of steps to achieve convergence, our numerical experiments show that in many instances the confidence region collapses to the exact ground state after few iterations.
A benchmark on a D-Wave 2000Q device shows how our chordal branch-and-bound (CBB) method can be used both to detect situations in which the quantum solution differs from the optimum and, more importantly, verify when the annealer has actually reached the ground-state energy and no further optimization is required. 

Our approach to verification consists of relaxing the initial optimization problem and, therefore, results in lower bounds to the ground-state energy, something impossible using standard variational techniques. Certification methods based on relaxations constitute a valid approach for benchmarking any heuristic optimisation, classical or quantum. The main purpose of this work is to demonstrate how these methods can be applied to asses the quality of the outputs of intermediate-scale quantum computing devices, whose timeliness is particularly motivated by recent progresses in the field~\cite{arute2019supremacy}.

\section{Preliminaries}
We consider classical spin systems whose configurations $\vec{\sigma} \coloneqq (\sigma_1,\dots,\sigma_N)$ are vectors of $N$ spin variables $\sigma_i \in \{-1,1\}$ to each of which a Hamiltonian $H$ assigns an energy $H(\vec{\sigma})$.
We are mostly interested in Hamiltonians of Ising type, that can be written in the form
\begin{equation}\label{eq:Hamiltonian}
  H(\vec{\sigma}) = \sum_{1 \leq i < j\leq N} J_{ij} \sigma_i \sigma_j + \sum_{i=1}^N h_i \sigma_i,
\end{equation}
with couplings $J_{ij}$ and local fields $h_i$.
The method we develop, however, is more general and can also be applied to Hamiltonians that are higher order polynomials of the spin variables $\sigma_i$ and couple three or more spins in a single term.
Among all the configurations of such a system there are those that achieve the minimal possible energy, also known as the ground-state energy and defined as
\begin{equation} \label{eq:ground_state_energy}
  E_g \coloneqq \min_{\vec\sigma \in \{-1,1\}^N} H(\vec\sigma) .
\end{equation}
This minimization is a polynomial optimization problem in the spin variables.
If the Hamiltonian is of the form in \eqref{eq:Hamiltonian}, then it is quadratic. For our purposes, solving the ground-state problem for a given Hamiltonian means finding $E_g$ and outputting a configuration that achieves it. Obviously, finding the ground state is an optimisation problem that can, in principle, be solved by brute force search.
This however quickly becomes infeasible as the number of configurations grows as $2^N$, restricting this approach to systems of few tens of particles.

Many spin models of interest in physics and beyond are characterized by a locality structure defined by a graph $G$: spins are located in the nodes of the graph and the interacting terms $J_{ij}$ are non-zero only between neighbouring sites, that is, spins connected by an edge of the graph. Local interactions also appear naturally in physical solvers of spin models, such as, for instance, a quantum annealer.
Such locality of interactions implies a sparsity of the resulting Hamiltonian and hence optimisation problem.
However, exploring this structure still remains a really non-trivial task:  
even for rather restricted classes of graphs, finding the ground state is an NP-hard problem.
This precludes the existence of any efficient and general algorithm.
The complexity of the Ising ground state problem thereby depends on subtle details of the problem class (see Appendix \ref{app:complexityoffindingIsinggroundstates} for more details).

It is convenient for what follows to present an equivalent formulation of the ground-state problem in which the energy is computed over the set of expectation values of products of spin variables, such as the $\sigma_i$ and $\sigma_i\sigma_j$ that appear in the Hamiltonian, instead of spin variables directly.
The connection with the original optimization problem \eqref{eq:ground_state_energy} is made by introducing the notion of a state of an $N$-spin system as a generic probability distribution $P$ over the set of configurations $\{-1,1\}^N$. For every function $f: \{-1,1\}^N \to \R$ we can then define its expectation value in the state $P$, denoted by 
\begin{equation}
  \langle f \rangle_P = \sum_{\vec\sigma \in \{-1,1\}^N} f(\vec\sigma)\, P(\vec\sigma) .
\end{equation}
In the following, whenever the connection to a specific physical state $P$ is not evident, we simply denote expectation values by $\langle f \rangle$. With this notation, the equivalent energy minimization problem reads
\begin{equation}\label{eq:ground_state_moments}
  H(\vec{\sigma}) = \min_P-\sum_{1 \leq i < j\leq N} J_{ij} \langle\sigma_i \sigma_j\rangle_P + \sum_{i=1}^N h_i \langle\sigma_i\rangle_P .
\end{equation}
We call any state $P$ that is supported only on the ground-state space (i.e., the collection of all configurations that achieve the ground-state energy) of a model a ground state and such $P$ are manifestly those that achieve the minimal possible expectation value for the Hamiltonian, i.e, $\min_P \langle H \rangle_P = E_g$.
Optimizing over probability distributions instead of spin configurations requires a similar exponential effort in the system size. Nonetheless, it opens the way for a relaxation in terms of an SDP problem, which is one of the main ingredients of our method, which we present in the following Section.

\section{The Chordal Branch and Bound algorithm}

In this section we describe the main tools we build upon to devise the CBB algorithm. First, we present the SDP relaxation we exploit to obtain both a lower and upper bound to the ground-state energy. These bounds define an energy confidence region in which the unknown ground-state energy provably lies. Then we introduce the branch-and-bound procedure as a tool to systematically improve this confidence region. Lastly, we describe the chordal extension method that exploits the sparsity of relevant physical models in order to increase the scalability of the SDP. Technical details of the algorithm and its implementation can be found in the Appendices~\ref{app:relaxationandchordal} and~\ref{app:CBBdetails}.

\subsection{SDP Relaxations}

As mentioned in the previous Section, the ground-state minimization problem can be relaxed to obtain an efficient method to derive lower bounds on the ground-state energy through the formulation in terms of expectation values~\eqref{eq:ground_state_moments}. Specifically we use a method pioneered by Lassere~\cite{lasserre2001explicit,lasserre2001global}, which relaxes the polynomial optimization over any distribution $P$ into an SDP.

Let us consider a vector $\vec x \coloneqq (x_\alpha)_{\alpha=1}^k$ of monomials of the spin variables.
For any state $P$ we define its moment matrix $\Gamma(P)$ with respect to $\vec x$ as the $k \times k$ matrix of expectation values $\Gamma_{\alpha\beta}(P) \coloneqq \langle x_\alpha x_\beta \rangle_P$.
Any such moment matrix $\Gamma(P)$, being defined via an outer product, is manifestly positive semidefinite, i.e., $\Gamma(P) \succeq 0$, and, depending on what the elements of $\vec x$ are, it further obeys certain linear constraints, which follow from the fact the monomials are made of spin variables. More precisely, the constraints reflect the two basic properties of these variables, namely that they take dichotomic values $\sigma_i \in \{-1,1\}$ and commute with each other.
This leads to conditions such as, for instance, $\langle \sigma_i \sigma_j \sigma_i \rangle_P = \langle \sigma_j \rangle_P$.

We illustrate this with an example: take as a generating set of monomials $\vec x$ the spin variables themselves together with the identity, i.e., $\vec{x} = \lbrace 1, \sigma_1, \ldots , \sigma_N \rbrace$. The resulting $\Gamma$ matrix takes the following form:
\begin{equation}\label{eq:example}
\Gamma =
\begin{pmatrix}
1 & \langle \sigma_1 \rangle & \langle \sigma_2 \rangle &
 \ldots & \langle \sigma_N \rangle  \\
\langle \sigma_1 \rangle & 1 &  \langle \sigma_1 \sigma_2  \rangle &  \ldots  & \langle \sigma_1 \sigma_N \rangle    \\
\langle \sigma_2 \rangle & \langle \sigma_1 \sigma_2  \rangle & 1 &
  \ldots  & \langle \sigma_2 \sigma_N \rangle  \\
\vdots & \vdots & & \ddots & \\
\langle \sigma_N  \rangle & \langle \sigma_1 \sigma_N \rangle & \ldots &  & 1  \\
\end{pmatrix}
\end{equation}
Notice how the expectation value of any Hamiltonian of the form given in \eqref{eq:Hamiltonian} can be expressed as a linear function of the entries of $\Gamma$, given by $\tr(h\,\Gamma)$, where $h$ is a matrix defined by the system Hamiltonian.
A lower bound to its ground-state energy can then be obtained by minimizing $\tr(h\,\Gamma)$ over all postive semidefinite matrices $\Gamma$ that fulfil the linear constraints discussed above, expressed also as linear functions of the entries of $\Gamma$ in terms of some matrices $F_m$,
\begin{eqnarray}\label{eq:relax}
  &\min_{\Gamma} & \tr(h\,\Gamma)  \nonumber\\
  &\text{s.t.}\qquad & \Gamma \succeq 0 ,  \\
  && \tr(F_m \, \Gamma) = 0 \, , \quad \forall \, m \in \{1,\dots,k\} \nonumber
  \label{eq:relax_end}
\end{eqnarray}
This defines an SDP relaxation of the problem, since not every such positive matrix $\Gamma$  satisfying the linear constraints encapsulated by the matrices $F_m$ necessarily arise as a moment matrix $\Gamma(P)$ of a physical state. In contrast with the original minimization, the presented SDP can be solved efficiently, since the amount of variables involved in the moment matrix scales only quadratically with the number of spins.

Interestingly, from the solution of the considered relaxations one can also extract a spin configuration with no additional computational cost. Let $\Gamma^\ast$ be the optimal solution to the SDP.
We can associate to it a configuration $\vec{\sigma}^\ast$ by taking the sign of the entries in that matrix that correspond to the expectation values $\langle \sigma_i \rangle$, namely set $\sigma_i^{\ast} \coloneqq \sign( \Gamma^{\ast}_{1,i+1} )$.
The energy of that configuration $H (\vec{\sigma}^\ast)$ clearly provides an upper bound to the ground-state energy.

Moreover, the approximation to the exact ground-state energy can be improved by considering the moment matrix generated by the vector $\vec{x}^{(\nu)}$ of all monomials of spin variables of degree up to $\nu$. For every such a vector, it is possible to construct the corresponding moment matrix $\Gamma^{(\nu)}$ and solve the corresponding relaxation, as in~\eqref{eq:relax}. This process defines a hierarchy of relaxations ordered according to the degree of the considered monomials $\nu$ that provides an asymptotically converging series of tighter and tighter lower bounds on the ground-state energy (see Appendix \ref{app:relaxationandchordal} for more details). All the steps in the hierarchy are efficient, as they define SDP problems involving matrices that scale polynomially with the number of spins.

\subsection{Branch and bound} 

The derived lower and upper bounds through the previous SDP relaxation can be combined with a so-called branch-and-bound (BB) technique to obtain a series of complementing bounds converging to the exact solution.
This is a general iteration strategy that has been applied in several different ways (see, for instance, Ref.~\cite{rendl2008solving} for a review).
The main ingredient of a BB iteration is the branching procedure, which consists in dividing the original problem into two sub-problems that correspond to the opposite cases of a dichotomic choice.
In the ground-state problem, it can be done by choosing a spin $i$ and considering the two subsets of spin configurations that have $\sigma_i  = \pm 1$ fixed.
Finding the ground state in both subcases can be cast as another ground-state problem for a modified graph where the vertex $i$ has been removed and the couplings have been modified accordingly.
Obviously, the value of the original ground-state energy is just the minimum between the solutions of the two sub-cases.

The trick is now to use the upper and lower bounds to reduce the number of branches to explore.
The BB procedure does that as follows:
\begin{enumerate}[(i)]
    \item start with the original graph and compute a lower and upper bound $z_{L},z_{U}$ to the ground-state energy, in our case using the previous SDP relaxation;
    \item if the bounds differ, choose a branching and compute lower and upper bounds for the two subcases;
    \item keep track of the best upper bound $\bar{z}_U$ encountered so far and discard all the explored branches in which the lower bound is higher than $\bar{z}_U$;
    \item from the reduced list of branches, pick the one corresponding to the lowest $z_L$, if it still differs from the best upper bound $\bar{z}_U$, go back to point (ii) and perform another branching;
    \item keep repeating until the lowest $z_L$ and the best upper bound $\bar{z}_U$ coincide.
\end{enumerate}

\begin{figure}[t]
  \centering
  \includegraphics[width=\linewidth]{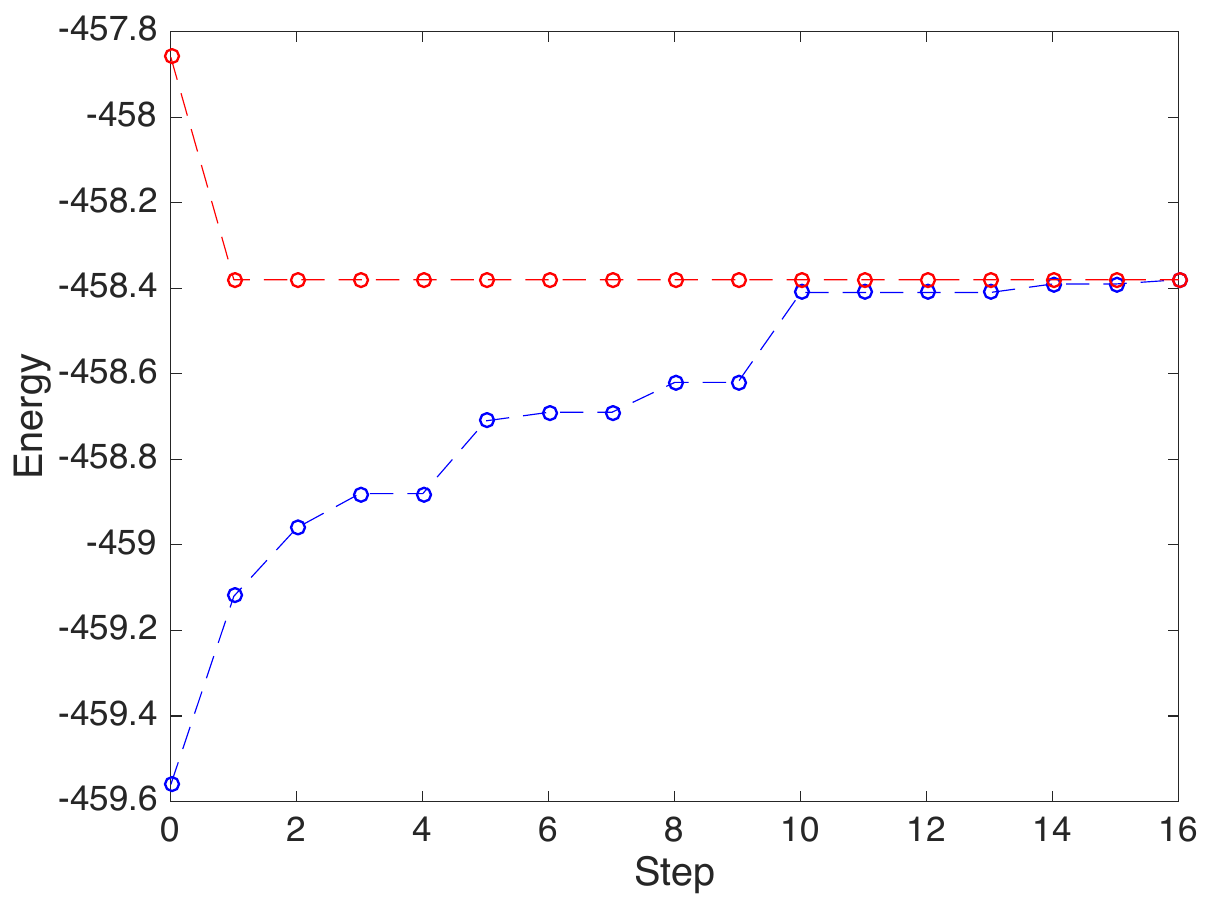}
  \caption{Example of sequences of upper and lower bounds obtained during the branching procedure with the chordal branch and bound method, for a 2D Ising model of lattice size $15\times15$.
    The convergence is met, yielding a certified value for the ground state energy and configuration, after less than $20$ steps, instead of having to explore all of the $2^{225}$ configurations.
  } 
  \label{fig:branch}
\end{figure}

Although the BB procedure always converges to the solution, it may require an exponential number of steps when implemented on hard instances. Yet, the method presents two important properties: (i) it provides a constantly improving energy range for the ground-state energy and (ii) at all steps, it is known whether the searched solution has been reached, possibly up to numerical precision, and no more steps are needed. In fact, we have observed that in many situations the algorithm can be stopped after a few steps because it has been able to find the solution. Fig.~\ref{fig:branch} shows a typical instance of the BB procedure in which the lower and upper bounds converge after a few iterations. More details about the implementation of the BB procedure can be found in Appendix~\ref{app:CBBdetails}.

\subsection{Chordal extension}

The last ingredient in our construction uses the fact that, for the considered spin systems with local interactions, the optimisation problem defined by the Hamiltonian \eqref{eq:Hamiltonian} is sparse.
As shown in Refs.~\cite{waki2006sums,lasserre2006convergent}, one can exploit this sparsity to derive a similar relaxation that is more scalable than the previous one.
Intuitively, the idea behind the modification is the following: for any pair of non-interacting spins $i,j$, the corresponding two-body expectation value $\langle \sigma_i \sigma_j \rangle$ is not needed for the computation of the energy.
Thus, a moment matrix with all two-body correlations is including some potentially unnecessary information.
Finding the minimal amount of moments that is sufficient to effectively constrain the optimisation of the energy helps defining a more efficient, and therefore scalable, relaxation.

\begin{figure}[t]
\centering
\includegraphics[width=0.9\linewidth]{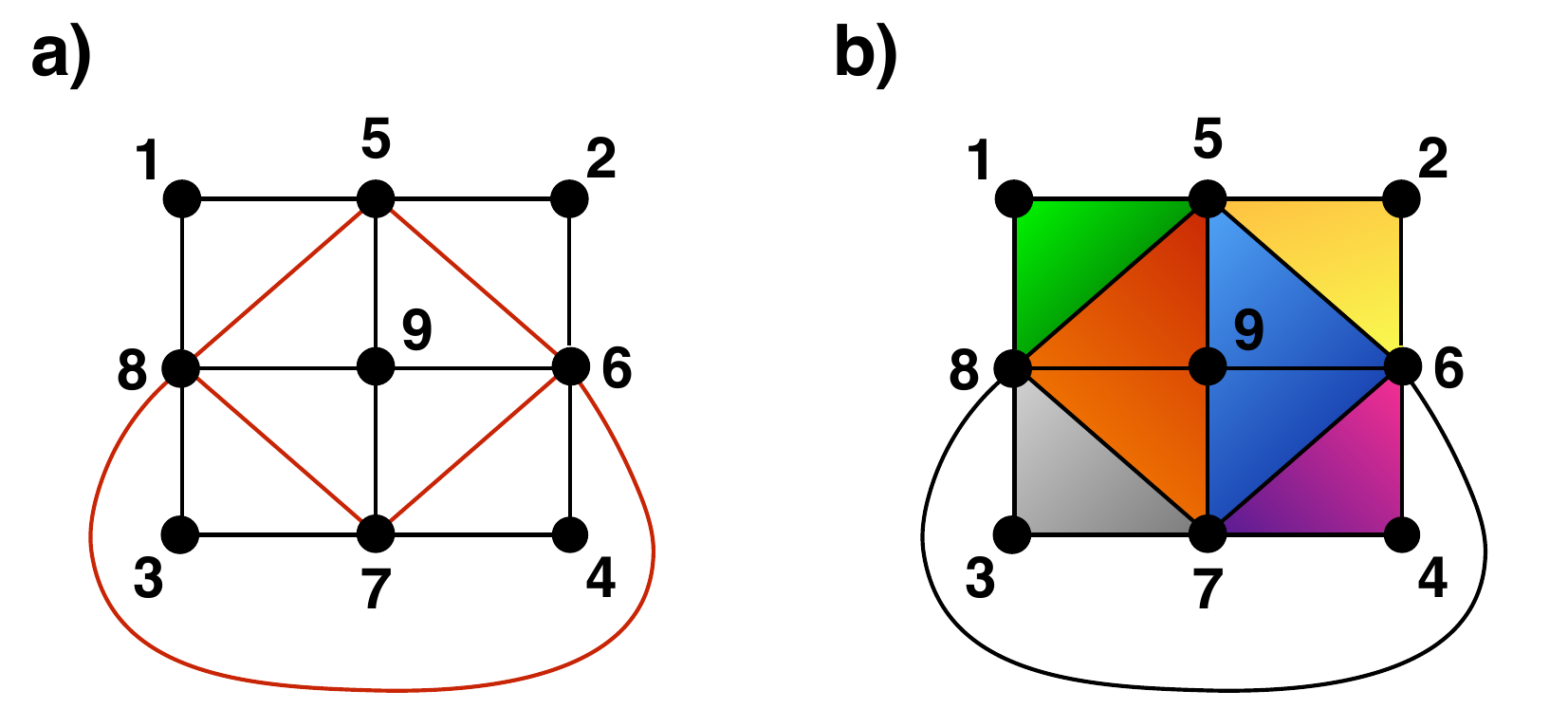}
\caption{a) An example of graph $G$ that is not chordal, together with an option of edges to add (shown in red) in order to make it chordal, obtaining a chordal extension $G_C$. The new edges are chosen to ``break'' the larger cycles, such as for example the cycle $\lbrace 1,5,9,8 \rbrace$.
  b) The graph $G_C$ and its corresponding maximal cliques $C_l$, shown in the different colours.
  In this example, the graph has $n_c = 6$ cliques, out of them two are composed of four vertices, while the remaining ones consist of three vertices each.}
\label{fig:chordal}
\end{figure}

To illustrate the method, let us suppose that the graph $\depgraphG$ of the problem is already chordal. A graph is said to be chordal if all its cycles of four or more vertices have a chord, i.e. an edge that is not part of the cycle but connects two vertices of the cycle.
If $G$ is not chordal, it is always possible to associate to it a so-called chordal extension $G_C$ by properly adding some edges (see Fig.~\ref{fig:chordal} for an example). Notice that the chordal extension of a graph is not unique, but a chordal extension can always be found, because the fully connected graph is chordal.
For a chordal extension to be useful for our CBB method it needs to be still relatively sparse, as in the example presented here.
In Appendix~\ref{app:relaxationandchordal} we provide a general poly-time technique to find good chordal extensions for all the cases studied in this paper. 
  
Once the chordal extension has been derived, one can then introduce a relaxation where the original $\Gamma$ matrix is replaced by a direct sum of smaller matrices $\Gamma_{l}$, constructed only from the spin variables belonging to the each of the $n_C$ maximal cliques, i.e. fully connected subgraphs of maximal size, of $G_C$.
Note that some spin variables appear in more than one clique, which means that the SDP does not completely decouple into separate SDPs for each block $\Gamma_{l}$, so the optimization involves moments appearing in multiple blocks.

As an illustration let us go back to the previous relaxation and suppose one wants to solve the 1D Ising model with Hamiltonian $H = \sum_{i =1}^N J_{i,i+1} \sigma_i \sigma_{i+1}$.
The corresponding dependency graph $G$ is already chordal and is composed of $N$ cliques $C_i = {\sigma_i, \sigma_{i+1}}$, with $i = 1,\ldots,N-1$.
Then, the matrix \eqref{eq:example} can be substituted by a direct sum of the $N$ blocks
\begin{align}
\Gamma_{C_i} &= \label{eq:example_block}
\begin{pmatrix}
1 & \langle \sigma_i \rangle & \langle \sigma_{i+1} \rangle  \\
 \langle \sigma_i \rangle & 1 &  \langle \sigma_i \sigma_{i+1} \rangle  \\
\langle \sigma_{i+1} \rangle & \langle \sigma_i \sigma_{i+1} \rangle &  1  \\
\end{pmatrix}.
\end{align}
Unnecessary expectation values in \eqref{eq:example}, such as $\langle \sigma_i \sigma_{i+2} \rangle$, no longer appear in any of the $N$ smaller blocks $\Gamma_{C_i}$, but all the expectation values that are needed to define the Hamiltonian as a linear function of the moment matrices are still present.
This simplification is particularly useful because it can significantly reduce the number of variables involved in the SDP and it reduces the size of the largest positive semidefinite block, which dramatically reduces computational footprint in the numerical algorithms solving the optimization.
In this example, we go from a matrix whose size increases quadratically with the number of spins, to a linearly increasing set of constant size matrices.
In general, the constraints between different blocks do not allow to split the problem into $n_C$ independent ones, but one can still see that the
scaling of the computational effort is dominated by the size of the largest block alone (see Appendix \ref{app:relaxationandchordal} for more 
details on the chordal extension hierarchy). 

As an illustration of the gain in scalability provided by the chordal extension, we compare the performance of the CBB method with a sparse Ising problem in the two cases of exploiting and not exploiting the chordal extension. As a benchmark of a sparse instance, we consider the standard 2D ferromagnetic Ising model in a statically disordered magnetic field (quenched disorder) that is picked independently from normal distributions of mean zero and variance $\sigma$ for each site. Similar results are obtained for other models with local interactions.
As a function of the disorder strength $\sigma$, the model undergoes a phase transition from a ferromagnetic ground state (in which all states are aligned with each other) to a disordered phase (in which, for extremely large disorder, the spins are aligned with the local magnetic fields).
For this model it is known that the ground state can in principle be found in polynomial time (see Appendix \ref{app:complexityoffindingIsinggroundstates}).

\begin{figure}[tb]
\centering
\includegraphics[width=0.47\textwidth]{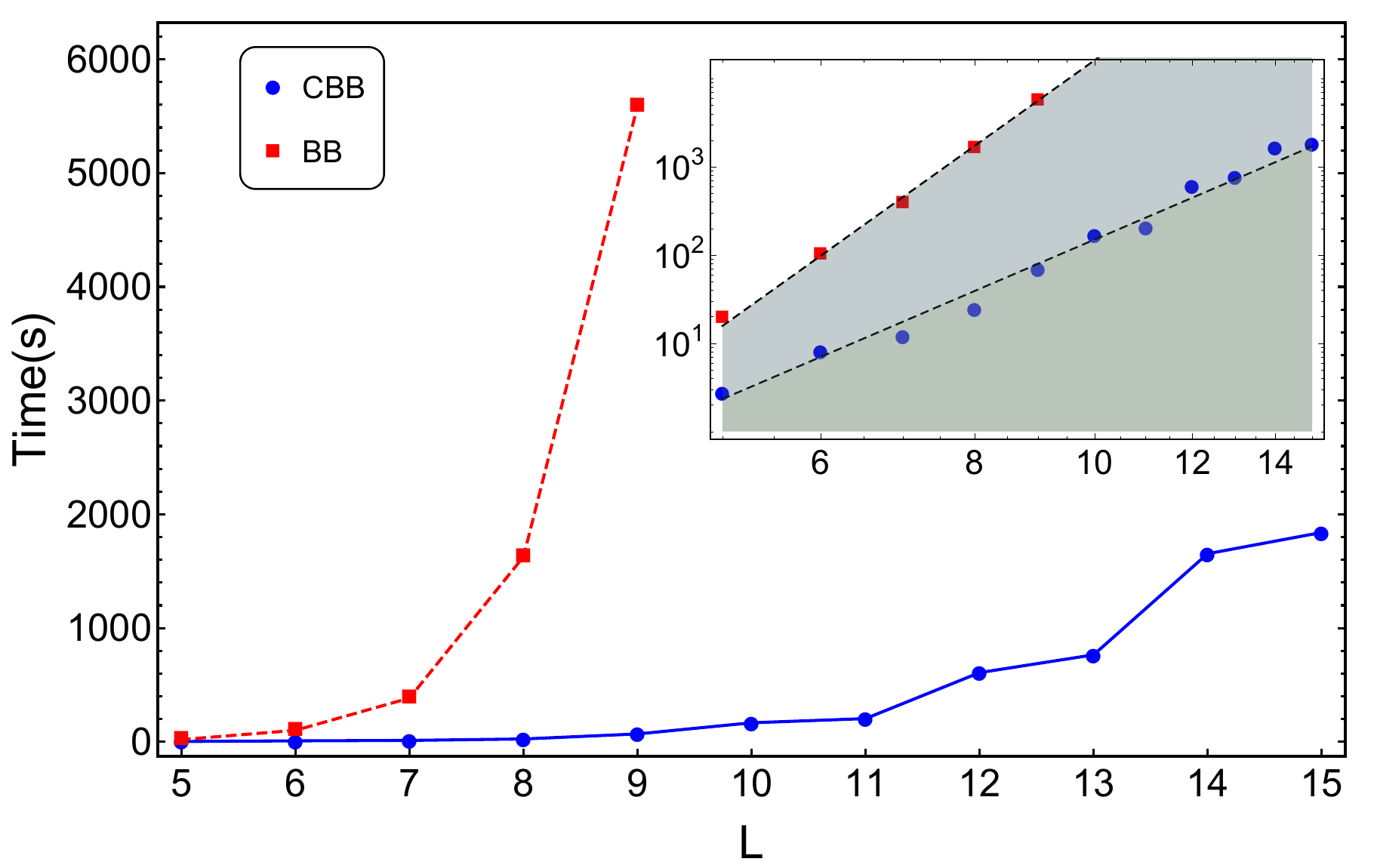}
\caption{Time comparison for a square lattice of $N$ sites between a standard SDP-based branch and bound (BB) algorithm and the one augmented with the chordal extension (CBB), as a function of the linear lattice size $L = \sqrt{N}$.
  The problem is finding the ground state energy for a 2D Ising ferromagnetic model with random Gaussian magnetic field, close to the phase transition at $\sigma = 1.5$, i.e. where the ground state is already partially disordered.
  The time estimation was averaged over $100$ disorder realizations, except for the largest system size, where the averaging was reduced to $10$ samples.
  Due to the large amount of disorder averaging, we limit ourselves to system sizes $L\leq 15$, far below the maximum sizes we can tackle on our hardware.
  The comparison is shown in a linear scale and double logarithmic in the inset.
  The dashed lines in the inset are power laws of the form $L^{10}$ and $L^6$, demonstrating the claimed polynomial scaling of the runtime $N^5$ for BB vs.\ $N^3$ for CBB.\label{fig:timeplot}}
\end{figure}

Indeed, the non-chordal BB method is able to find the solution with an effort that scales roughly with a $N^5$ dependence, see Figure~\ref{fig:timeplot}. However, and especially in the interesting region close to the phase transition, fast growing memory requirements and runtime make the method impractical for systems that are larger than $15 \times 15$ on the hardware we have at our disposal.
The CBB method, in contrast, allows us to solve systems of over $35 \times 35$ sites on the same hardware, due to both lower memory requirements and a very significantly reduced runtime, both in terms of absolute numbers and in terms of scaling (see Figure~\ref{fig:timeplot} for a comparison). 
When using the chordal extension, the method scales roughly as $N^3$, as opposed to the $N^5$ dependence without it.

\section{Verifying the solution of a quantum annealer}
Once all the ingredients of the method have been presented, we now turn to the main part of our work and show how to apply our approach to verify the results of energy minimizations performed on an actual quantum annealing device.

Numerical computations in this work were run on a workstation with an Intel Xeon E5-1650v4 processor with six physical cores clocked at 3.60~GHz base frequency and 128~GByte RAM.
Due to the polynomial scaling of the method, much larger system sizes can be reached with more powerful hardware.
The sparse semidefinite relaxations were generated by Ncpol2sdpa~\cite{wittek2015ncpol2sdpa}, and the semidefinite programs were solved by Mosek~\cite{mosek2010mosek}.
The code for the experiments is available under an open source license~\footnote{\url{https://gitlab.com/FBaccari/Chordal_BB}}.

%

%

\subsection{Verifying solutions for a triangular graph}

To show the flexibility of the CBB method and also to verify a quantum annealing solution for the largest system size simulable on a state-of-the art annealer, we considered a 2D triangular lattice.
In fact, spin models on the triangular lattices display a wealth of interesting physical phenomena, many driven by the possibility to have frustrated interactions.
To remain in a regime that is comparable to the benchmarking we did before, we however concentrate on the interplay of ferromagnetic interactions with a disordered magnetic field (for a numerical analysis of the corresponding phase transition, see Appendix \ref{app:phasetrans}).

We used a D-Wave 2000Q quantum annealer with 2040 functional qubits.
The chip had 8 faulty qubits and the corresponding couplings were removed from a full-yield $16\times 16$ Chimera graph.
We used the virtual full-yield Chimera graph abstraction to ensure consistent embeddings and improve the quality of the results.
The coupling strengths were automatically scaled to the interval $[-1, 1]$, and the logical qubits used a coupling strength of $-2$ to hold a chain of physical spins together.
The minor embedding was a heuristic method, yielding a chain length of $7$.
We also tried chains up to length $22$, without significant change in the results, showing that the scaling in the couplings ensures that the chains do not break.
For each data point, we sampled a thousand data points and chose the one with the lowest energy as the optimum.
This takes constant time irrespective of the values set, in the range of milliseconds.
The flux bias of the qubits was not offset.

\begin{figure*}[t!!]
\centering
\includegraphics[width=0.7\textwidth]{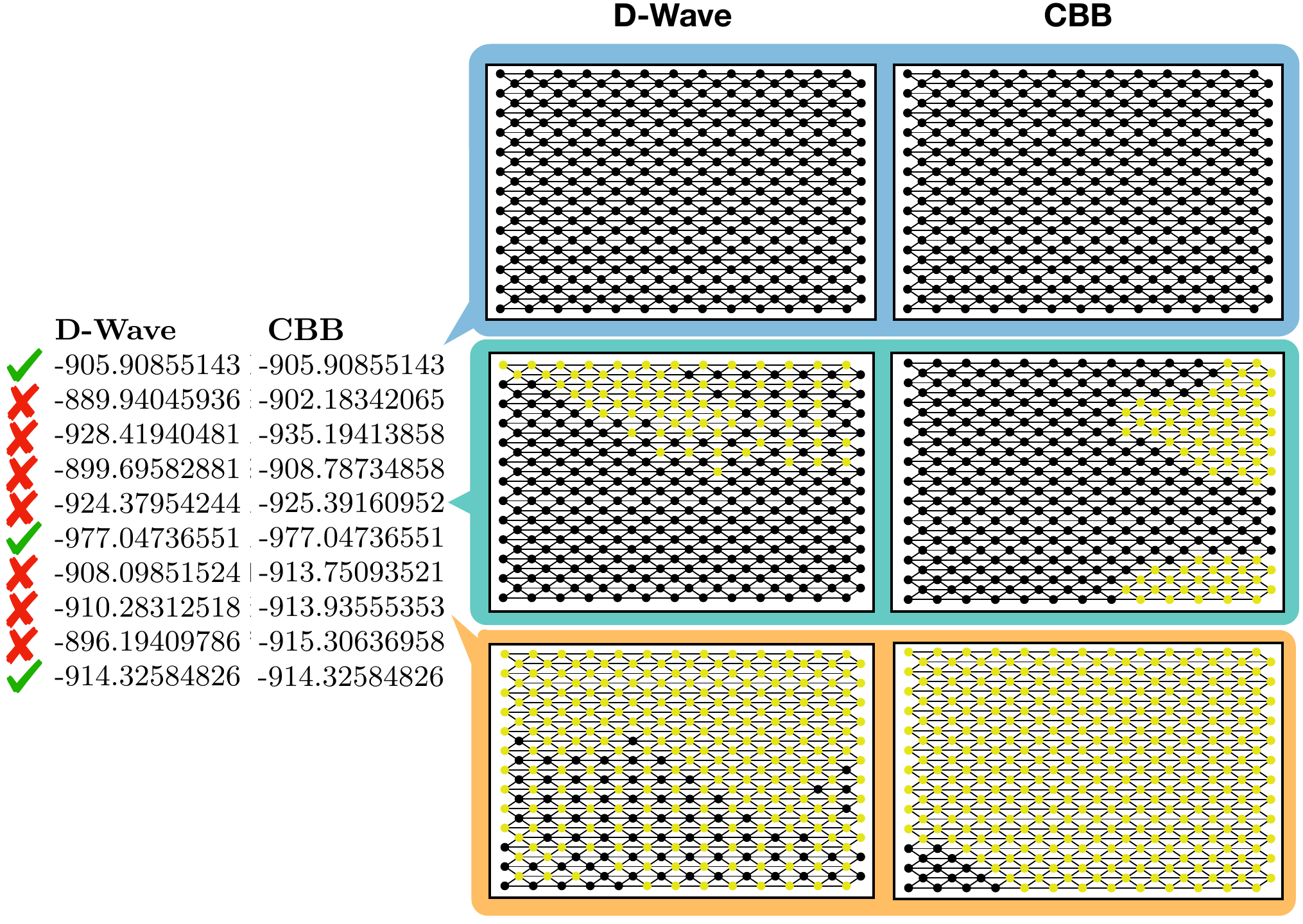}
\caption{Left: Comparison of the lowest energy value for a disordered phase at $\sigma=1.5$ on a triangular lattice on a 2040-qubit D-Wave quantum annealer and the chordal branch and bound algorithm. Notice that in all the instances considered here the confidence region provided by CBB converged in few steps and the algorithm returned the exact ground state energy. 
Right: comparison of the corresponding ground state spin configuration both for a case where D-Wave achieves the lowest energy and for two cases where it does not.
Yellow spins are $+1$ and black one $-1$.
It can be seen clearly that, even when the energies provided by the two methods are close, the corresponding spin configurations can be very different.
This shows that the excited state that the D-Wave quantum annealer returns, does not necessarily resemble the globally optimal solution.}
\label{fig:dwaveresults}
\end{figure*}

Both the quantum annealer and the CBB simulation were done for the same disorder realizations (the disorder in the annealer is fully programmable) to obtain directly comparable results.
We observe that there are indeed some cases in which even after 1000 repetitions, the lowest energy found by the quantum annealer is still higher than the exact value computed by CBB.
Here, the optimal spin configuration found by the quantum annealer typically differs markedly from the the output of our method, which detects that the quantum device probably got stuck in a local minimum. Interestingly, our method is also able to certify that, for some disorder realizations, the quantum annealer is able to find the exact ground-state energy. It does that typically in a very short time. This is true even for intermediate disorder strengths, around $\sigma=1.5$, where the ground-state spin pattern shows macroscopic islands of aligned spins whose precise shape and positions depend non-trivially on the disorder realization, see Fig.~\ref{fig:dwaveresults}.
Verifying that the annealer did reach the correct solution is impossible with standard variational approaches used so far and clearly demonstrates the relevance of the introduced CBB method for the benchmarking of quantum optimizers.

\subsection{Towards the verification for a Chimera graph}

Lastly, we consider the application of CBB to a denser graph. For this purpose we choose the Chimera architecture~\cite{bunyk2014architectural}, which is the natural graph on the D-wave 2000Q hardware.
The corresponding graph is composed of $K_{4,4}$ fully connected bipartite unit cells, consisting of 8 spins – 4 horizontal and 4 vertical – with edges between each horizontal/vertical pair. These unit cells are arranged to form a 2D square lattice of size $L$ and a total number of $N = 8\,L^2$ spins.
Because of the in-cell connectivity, such a graph is clearly non-planar and thus has the potential to encode NP-hard Ising models.
Even though the Chimera graph is non-planar and denser than 2D rectangular and triangular lattice, using the chordal extension still gives a remarkable advantage, allowing us to reach system sizes of $L=9$, compared to just $L=5$ (on the same hardware) for an SDP-based BB method without the chordal extension.

Although the D-Wave 200Q quantum annealer is currently implementing a Chimera graph with $2040$ functional physical qubits, they are seldom actually used as logical spins.
Most recent studies encode each $K_{4,4}$ cell as a single logical spin, in order to suppress errors due to the finite size and qubit quality of the system~\cite{mandra2018deceptive}.
This results in effectively solving Ising models on a 2D square lattice which, being planar, is actually proven to be polynomially solvable (cfr. Appendix \ref{app:complexityoffindingIsinggroundstates}). 
The numerical test was performed on the actual Chimera graph.
This opens up the way to benchmarking future annealing devices, once their physical qubit quality has improved to a point that makes the individual spins useful, in the much more interesting regime of non-planar graphs.

\section{Comparison with other verification methods}

The problem of verifying a quantum optimizer is very rich and has many interesting sides. We can identify two relevant players in the problem: the provider and the user. A first problem consists of verifying that the quantum hardware performs some form of quantum process, say quantum annealing, with no classical analogue or that is hard to simulate classically. This type of verification is available to the provider when constructing and testing the device, but generally impossible for the user. Here, the verification methods must be quantum specific. The provider may tomographically reconstruct the different quantum steps in the optimization, or may simply want to certify that the device, when solving the optimization problem, generates a large amount of some given quantum properties, such as coherence, entanglement or quantum non-locality. This may be a direct evidence of some form of ``quantumness", but it does not guarantee that the device performs better than a classical approach, as at the moment it is not clear which quantum properties -- if any -- could provide a quantum computational advantage. Moreover, it is difficult to see how this approach could detect instances where the quantum device gets stuck in a local minimum, the typical challenge in the considered optimization problems.

Our work falls into a second class of methods, which attempts at verifying the device only from the outputs it produces, without any reference to the quantum process that gave raise to it. Note that, being based only on a classical output, none of these methods is quantum specific and all can be applied to any optimizer, classical or quantum. This is the type of verification that is more relevant to the user. For optimization problems, this verification will mostly be based on heuristics: most of the relevant problems are NP-hard and it is expected that will remain hard also for quantum computers. Yet, one can not exclude the possibility that quantum devices might eventually give better solutions than any classical solution ``on average", or at least for some families of practically relevant problems. In fact, the search for a quantum advantage is one of the main focus of research on quantum algorithms for classical optimisation problems. One can identify three ways of verifying the outputs of quantum optimizers (or, as said, any optimizer).

\emph{Planted solutions:} in this approach, an optimizer is run on problems for which the solution is known~\cite{hen2015probing,katzgraber}. This is a way of testing that the considered algorithm is able to obtain the expected solution and hope that it will perform equally well for other problems for which the solution is unknown. In our opinion, this approach is especially relevant for the development of quantum optimizers, for instance to understand which instances are difficult for them. But it also has clear limitations. First, the considered optimization problems are so diverse that it is conceivable that a device may not be able to find the exact solution for problems where it is known in advance while giving reasonably good results for other problems of relevance.
Second, testing a device on problems for which the solution is known can not lead to any quantum advantage by definition. Therefore, if any quantum advantage were to be demonstrated, it will never be by running a quantum device on problems with a planted solution.

\emph{Variational methods:} we group in this class any approach providing a candidate solution, not necessarily the optimal one, to the problem. By definition, these approaches provide upper bounds to the searched solution, as the quantum optimizer. Simulated annealing~\cite{Kirkpatrick1983} or tensor-network algorithms~\cite{TNOrus} are notable classical examples of this approach. If the value provided by the quantum optimizer $E_q$, is larger than the best value obtained by one of these classical approaches, $E_c$, one can certify that the quantum device has not reached the searched solution, $E_g$. Classical variational methods have a long tradition and can deal with very large systems, at the moment larger than those available for quantum optimizers. If, however, quantum optimizers will ever reach the quantum advantage, one will encounter a situation in which they will be able to provide the smallest available upper bound to the solution. In a resulting scenario where $E_g\leq E_q<E_c$, the limitations of classical variational methods are clear: it is hard to determine whether the better performances of quantum optimizers are an indication of to their intrinsically higher computational power or the lack of a more efficient classical optimization algorithm.

\emph{Relaxations:} this is the approach considered in this work and, more in general, we refer here to any method providing a lower bound to the searched ground-state energy (note however that our approach can also be used to provide an upper bound with the same computational effort). These methods have much less history and cannot reach, at the moment, problem sizes comparable to variational methods. However, and because of their complementarity, they have clear merits. As shown by our demonstration using the D-Wave machine, these methods can give a termination criterion for  any optimisation heuristics, certifying that the searched solution has been reached and no more rounds are needed.  But even when a gap remains, the lower bounds obtained through relaxations can be complemented with the best upper bound, being quantum or classical, and provide an energy range in which the solution provably lies. Clearly, no quantum advantage is possible for those problems where the obtained lower bound matches the solution obtained by a classical optimiser. On the contrary, problems where there is a gap between the best known classical solution and the lower bound may be good candidates when searching for a quantum advantage. Finally, if the bound only matches the output produced by a quantum device, no classical approach will ever provide a strictly better solution. In our opinion these properties make this approach valuable and, in particular, especially relevant to advance the study and search for a quantum advantage in classical optimization problems.

\section{Conclusions}
%
We introduced the chordal branch-and-bound (CBB) method that uses a hierarchy of efficiently computable upper and lower bounds on the ground-state energy of classical spin systems and exploits the physical locality structure of relevant Hamiltonians. 
Our numerical results show that the iterative branch-and-bound process often converges, providing an exact and certified value for the ground-state energy, together with a ground-state configuration. Even for those cases where convergence is not met, the method always provides with a polynomial effort an energy range where the searched solution provably lies. Moreover, the obtained lower bound can be used to benchmark the output of both quantum and classical optimization methods. We focused here on their use to assess the quality of current quantum annealers beyond their comparison to classical variational approaches. In particular, we were able to identify instances were the quantum optimizer reached the actual ground state energy without resorting to planted solution problems. To our knowledge, to date this is the only approach providing this type of certification.

It would definitely be interesting to explore the performance of our method on other relevant Ising models that are proven to be NP-hard.
The complexity of the model might result in an exponential number of branch-and-bound steps to achieve convergence to the ground-state energy. However, let us stress that NP-hardness is a worst-case feature, hence it might be the case that an efficient convergence to the ground states is achievable on average even for some NP-hard models.
In this respect, CBB can be essential to provide numerical evidence that identifies which Ising models are uniformly hard, among the NP-hard ones. 

From a general viewpoint, recent progress of quantum computing devices \cite{arute2019supremacy} urge for the development of methods to benchmark them. Our work is an example of such effort and we are confident that the approach adopted here for benchmarking, based on relaxations of the initial problem, may find applications in other similar contexts.



\section{Acknowledgements}

A.A., F.B. and C.G. thank P.W. for many insightful discussions and his passion for research. We also acknowledge useful discussions with J. Tura and M. Navascu\'{e}s. This work is supported by the Spanish Ministry MINECO (QIBEQI FIS2016-80773-P and TRANQI PID2019-106888GB-I00, Severo Ochoa SEV-2015-0522 and PhD fellowship), the European Union's Marie Sk\l{}odowska-Curie Individual Fellowships (IF-EF) programme under GA: 700140, the Generalitat de Catalunya (CERCA Program, QuantumCAT and SGR 1381), Fundacio Privada Cellex and Mir-Puig, computational resources granted by the High Performance Computing Center North (SNIC 2015/1-162 and SNIC 2016/1-320), and a hardware donation by Nvidia, the Perimeter Institute for Theoretical Physics, the Government of Canada through Industry Canada and the Province of Ontario through the Ministry of Economic Development and Innovation, the ERC CoG QITBOX, AdG CERQUTE and the AXA Chair in Quantum Information Science.

\bibliography{bibliography}


\appendix

\section{Complexity of finding Ising ground states}
\label{app:complexityoffindingIsinggroundstates}
There is a wealth of results on the worst-case complexity of finding the ground state of various Ising models~\cite{Grotschel1987polyhedral}.
Thereby ``worst-case complexity'' is the complexity of the hardest instances within a class of families of problems of increasing size.
How hard it is to solve the ground-state problem of such a family varies with the interaction graph and can crucially depend on seemingly unimportant details.
We consider Hamiltonians that are polynomials (with fixed finite degree and finite precision coefficients) in the spin variables in the form
\begin{equation}\label{eq:Hamiltonian1}
  H(\vec{\sigma}) = \sum_{1 \leq i < j\leq N} J_{ij} \sigma_i \sigma_j + \sum_{i=1}^N h_i \sigma_i,
\end{equation}
with couplings $J_{ij}$ and local fields $h_i$, and interaction \mbox{(hyper-)}graph $\interactiongraphG \coloneqq (\interactiongraphVset,\interactiongraphEset)$.
The size of a problem is the number of vertices $N \coloneqq |\interactiongraphVset|$.
We say that an Ising model of the form \eqref{eq:Hamiltonian1} has no fields if $h_i = 0$ for all $i$, we say it has an external field if all $h_i = h$ for all $i$ and some $h$, and we say it has on-site fields if all $h_i$ can be chosen independently.

Finding the ground state of Hamiltonians of the form \eqref{eq:Hamiltonian1} for arbitrary graphs is in general NP-hard~\cite{Barahona1982complexity}, even without any fields.
This is still true for $J_{i,j} \in \{-1,0,1\}$ and $\interactiongraphG$ a finite 3D cubic grid graph and even for $\interactiongraphG$ a cubic two-layer 2D grid~\cite{Barahona1982complexity}.
In contrast, for planar graphs $\interactiongraphG$ and without local fields, the ground state can be found efficiently even without the restriction $J_{i,j} \in \{-1,0,1\}$~\cite{Grotschel1987polyhedral}.
With the restriction $J_{i,j} \in \{-1,0,1\}$ this even holds for toroidal graphs (grids on a torus, i.e., systems with periodic boundary conditions)~\cite{Grotschel1987polyhedral}.
Similarly, if $J_{i,j} \geq 0$, then even some systems with local fields can be solved in polynomial time~\cite{Grotschel1987polyhedral}.
On the other hand, for general planar graphs with interactions $J_{i,j} = 1$ and uniform external field $h_i = 1$, finding the ground state is again NP-hard~\cite{Barahona1982complexity}.
Ref.~\cite{Grotschel1987polyhedral} contains a list of further concrete models whose complexity is either known to be in P or proven to be NP-hard.

These hardness results are typically obtained by reducing the ground state problem to the so-called max-cut problem, which is known to the NP-hard.
The polynomial time algorithms to solve the other families of systems, in turn, work by finding perfect matchings~\cite{Barahona1982complexity} or rely on so-called max-flow/min-cut methods~\cite{Grotschel1987polyhedral}.
\section{Ingredients of the chordal branch-and-bound method}
\label{app:relaxationandchordal}
In this Appendix we describe in detail the Lasserre hierarchy and the chordal extension method used to derive lower bounds to the ground-state energy of a classical Hamiltonian.
To make this part self-contained, we revise the general method by using the notation use in the main text, while making comparisons to the more typical framework of polynomial optimisation.
Moreover, for the sake of completeness we will repeat some notions that have already been introduced in the main text.

We begin by recalling that a state of an $N$ spin system is a probability distribution $P$ over the set of configurations $\{-1,1\}^N$. Expectation values of a function $f$ for the state $P$ are denoted by $\langle f \rangle_P$, or even shorter by $\langle f \rangle$ if the connection to a specific states is not evident.
The ground state is defined by any state $P$ achieving the minimal possible expectation value for the Hamiltonian, i.e, $\min_P \langle H \rangle_P = E_g$, that is,
\begin{equation}\label{eq:ground_state_moments_app}
 H(\vec{\sigma}) = \min_P-\sum_{1 \leq i < j\leq N} J_{ij} \langle\sigma_i \sigma_j\rangle_P + \sum_{i=1}^N h_i \langle\sigma_i\rangle_P .
\end{equation}

\subsection{The hierarchy of lower bounds}

Let $\vec{x}^{(\nu)}$ be the vector of all monomials of spin variables of degree up to $\nu$.
For such vector and state $P$ we define its moment matrix $\Gamma^{(\nu)}(P)$ as the $k \times k$ matrix of expectation values $\Gamma^{(\nu)}_{\alpha\beta}(P) \coloneqq \langle x_\alpha^{(\nu)}\,x_\beta^{(\nu)} \rangle_P$.
It is not hard to see that for any state $P$, any moment matrix $\Gamma^{(\nu)}(P)$ is positive semidefinite, i.e., $\Gamma^{(\nu)}(P) \succeq 0$, and that, depending on what the elements of $\vec x^{(\nu)}$ are, it further obeys certain linear constraints that reflect the two basic properties of classical spin variables of taking dicotomic values $\sigma_i \in \{-1,1\}$ and commuting with each other.
These properties imply relations among different expectation values such as, for example, $\langle \sigma_i \sigma_j \sigma_i \rangle_P = \langle \sigma_j \rangle_P$, which can be expressed in terms of some matrices $F_m $ such as $\tr(F_m \Gamma^{(\nu)}(P))=0$.

For a sufficiently large value of $\nu$, the corresponding moment matrix $\Gamma^{(\nu)}$ contains all the expectation values needed for the computation of the energy, in the sense that there is a matrix $h$ (depending on the $J_{ij}$ and $h_i$ in case the Hamiltonian is of the form in \eqref{eq:Hamiltonian1}), such that for any physical state $P$ it holds that $\langle H \rangle_P = \tr(h\,\Gamma^{(\nu)}(P))$.
If this is the case, for that given $\nu$, one can relax the ground-state problem by minimising the energy over all matrices $\Gamma^{(\nu)}$ that are positive semidefinite and fulfil the above mentioned linear constraints, rather than over those that can actually arise from a physical state $P$. The resulting optimisation problem reads
\begin{align}\label{eq:relaxapp}
  E_g^{(\nu)} = \min_{\Gamma^{(\nu)}}  \tr(h\,\Gamma^{(\nu)}) & \nonumber \\
  \text{s.t.}\qquad \Gamma^{(\nu)} &\succeq 0 ,  \\
  \forall m \in \{1,\dots,k\}\colon \tr(F_m \, \Gamma^{(\nu)}) &= 0 . \nonumber
\end{align}
The solution to this minimisation problem $E_g^{(\nu)}$ provides a lower bound on the true ground-state energy, i.e., $E_g^{(\nu)} \leq E_g$. This follows from the fact that the set of all matrices $\Gamma^{(\nu)}$ that are positive semidefinite and fulfil the above mentioned linear constraints is larger than the set of moment matrices that can actually arise from a physical state $P$.
It is further obvious that the $E_g^{(\nu)}$ values are ordered in the sense that $E_g^{(\nu)} \leq E_g^{(\nu+1)}$ for any $\nu$. Remarkably, if all the relevant $F_m$ are taken into account, the bounds actually converge to the true ground-state energy for any fixed finite system size and Hamiltonian $H$, in the sense that $\lim_{\nu\to\infty} E_g^{(\nu)} = E_g$~\cite{lasserre2001global,pironio2010convergent}. This does not mean that convergence can only be attained in the limit. As a matter of fact, there are situations in which convergence is attained at a finite value of $\nu$.

To illustrate the first levels of the hierarchy, let us go back to the example presented in the main text with moment matrix
\begin{equation}\label{eq:example2}
\Gamma =
\begin{pmatrix}
1 & \langle \sigma_1 \rangle & \langle \sigma_2 \rangle &
 \ldots & \langle \sigma_N \rangle  \\
\langle \sigma_1 \rangle & 1 &  \langle \sigma_1 \sigma_2  \rangle &  \ldots  & \langle \sigma_1 \sigma_N \rangle    \\
\langle \sigma_2 \rangle & \langle \sigma_1 \sigma_2  \rangle & 1 &
  \ldots  & \langle \sigma_2 \sigma_N \rangle  \\
\vdots & \vdots & & \ddots & \\
\langle \sigma_N  \rangle & \langle \sigma_1 \sigma_N \rangle & \ldots &  & 1  \\
\end{pmatrix} .
\end{equation}
It is clear now that that moment matrix represents the hierarchy \eqref{eq:relaxapp} at level $\nu =1$. If we take the special case of $N = 3$ spins, the whole moment matrix reads
\begin{equation}\label{eq:examplelv1}
\Gamma^{(1)} = \left(
\begin{array}{cccc}
1 & \langle \sigma_1 \rangle & \langle \sigma_2 \rangle & \langle \sigma_3 \rangle  \\
\langle \sigma_1 \rangle & 1 &  \langle \sigma_1 \sigma_2  \rangle & \langle \sigma_1 \sigma_3 \rangle    \\
\langle \sigma_2 \rangle & \langle \sigma_1 \sigma_2  \rangle & 1  & \langle \sigma_2 \sigma_3 \rangle  \\
\langle \sigma_3  \rangle & \langle \sigma_1 \sigma_3 \rangle & \langle \sigma_1 \sigma_3 \rangle & 1  \\
\end{array}
\right)
\end{equation}
Going at level at level $\nu = 2$ for a system of $N = 3$ spins, the corresponding moment matrix take the following form:
\begin{widetext}
\begin{equation}\label{eq:examplelv2}
\Gamma^{(2)} = \left(
\begin{array}{ccccccc}
1 & \langle \sigma_1 \rangle & \langle \sigma_2 \rangle &
\langle \sigma_3 \rangle & \langle \sigma_1  \sigma_2 \rangle &
\langle \sigma_1  \sigma_3 \rangle & \langle \sigma_2  \sigma_3 \rangle \\
\langle \sigma_1 \rangle & 1 &  \langle \sigma_1 \sigma_2  \rangle & \langle \sigma_1  \sigma_3 \rangle & \langle \sigma_2 \rangle & \langle \sigma_3 \rangle & \langle \sigma_1  \sigma_2 \sigma_3 \rangle  \\
\langle \sigma_2 \rangle & \langle \sigma_1 \sigma_2  \rangle & 1 &
\langle \sigma_2 \sigma_3  \rangle & \langle \sigma_1 \rangle &
\langle \sigma_1  \sigma_2 \sigma_3 \rangle & \langle \sigma_3 \rangle \\
\langle \sigma_3 \rangle & \langle \sigma_1 \sigma_3  \rangle &
\langle \sigma_2 \sigma_3  \rangle & 1 & \langle \sigma_1  \sigma_2 \sigma_3 \rangle & \langle \sigma_1 \rangle & \langle \sigma_2 \rangle\\
\langle \sigma_1 \sigma_2  \rangle & \langle \sigma_2  \rangle & \langle \sigma_2  \rangle & \langle \sigma_1  \sigma_2 \sigma_3 \rangle & 1 & \langle \sigma_2 \sigma_3  \rangle & \langle \sigma_1 \sigma_3  \rangle \\
\langle \sigma_1 \sigma_3  \rangle & \langle \sigma_3  \rangle &  \langle \sigma_1  \sigma_2 \sigma_3 \rangle & \langle \sigma_1  \rangle &\langle \sigma_2 \sigma_3  \rangle & 1 &\langle \sigma_1 \sigma_2  \rangle \\
\langle \sigma_2 \sigma_3  \rangle & \langle \sigma_1  \sigma_2 \sigma_3 \rangle & \langle \sigma_3  \rangle & \langle \sigma_2  \rangle & \langle \sigma_1 \sigma_3  \rangle & \langle \sigma_1 \sigma_2  \rangle & 1\\
\end{array}
\right)
\end{equation}
\end{widetext}

Matrix $\Gamma^{(1)}$ is contained in $\Gamma^{(2)}$ as a principal minor -- the one generated by the first 4 rows and columns. This makes it so that the second level directly implies the constraint implied by the first one, while the non-negativity of a bigger moment matrix results in a generally more stringent test.
Moreover, the expectation value of any Hamiltonian of the form given in \eqref{eq:Hamiltonian1} can be expressed as a function of the entries of both moment matrices as $\tr(h\,\Gamma^{(\nu)}) = J_{12} \Gamma^{(\nu)}_{15} + J_{13} \Gamma^{(\nu)}_{16} + J_{23} \Gamma^{(\nu)}_{17} + h_1 \Gamma^{(\nu)}_{12} + h_2 \Gamma^{(\nu)}_{13} + h_3 \Gamma^{(\nu)}_{14}$ for $\nu = 1,2$.
Similarly, one can see how the linear constraints $F_m$ on the entries reflect the properties of the spin variables.
Dichotomy directly imposes the conditions $\Gamma^{(\nu)}_{ii} = 1$ on the diagonal variables and, combined with commutations, allows to identify some of the entries, such as for example $\Gamma^{(1)}_{12} = \Gamma^{(1)}_{21}$ and $\Gamma^{(2)}_{13} = \Gamma^{(2)}_{25}$.

As a comparison with previous similar methods, notice that the branch-and-bound technique introduced in Ref.~\cite{rendl2008solving} exploits a lower bound method that is almost equivalent to the first level of the relaxation \eqref{eq:relaxapp}, with the addition of some hand crafted linear constraints.
Indeed, the mentioned relaxation can be obtained by considering a moment matrix generated by the set of monomials composed of the spin variables only, without the identity operator. Hence, it results in the first level of the Lasserre hierarchy, diminished by the absence of the one (the first) row of the matrix.
In contrast, the hierarchy discussed here allows to systematically construct an infinite family of increasingly precise relaxations that yield better and better bounds at the price of an increasing computational cost.

\subsection{Exploiting sparsity via the chordal extension}
Depending on the kind of system considered, the optimisation problem defined by the Hamiltonian \eqref{eq:Hamiltonian1} can be sparse.
As is shown in Refs.~\cite{waki2006sums,lasserre2006convergent}, we can exploit this sparsity to derive a more scalable relaxation than \eqref{eq:relaxapp}.

The method works as follows: take the dependency graph $\depgraphG$ of the problem and check if it is chordal.
As mentioned in the main text, a graph is said to be chordal if all its cycles of four or more vertices have a chord, i.e. an edge that is not part of the cycle but connects two vertices of the cycle.
If $G$ is not chordal, construct a so-called chordal extension $\depgraphG_C$ of $G$ by suitably adding edges until the graph is chordal.
The chordal extension of a graph is not unique, but a chordal extension can always be found, simply because any fully connected graph is chordal.
As it will became clear in the following, for our purposes it is crucial to find a chordal extension that is still relatively sparse.
A poly-time method that works well in this respect for all the cases studied here is to compute an approximate minimum degree ordering of the graph nodes, followed by Cholesky factorization of a positive semidefinite matrix with the associated sparsity pattern~\cite{vandenberghe2015chordal}.
Once a specific chordal extension $\depgraphG_C$ is constructed, it will contain a number of $n_C$ maximal cliques $C_l \subset \interactiongraphVset$.
A clique, that is a fully connected subgraph, is maximal if it cannot be extended by including any other adjacent vertex.
Since the graph $\depgraphG$ represents a sparse Hamiltonian, and $\depgraphG_C$ is obtained from $\depgraphG$ by simply adding some edges, the function \eqref{eq:Hamiltonian1} can be decomposed into a sum $H = \sum_l H_{C_l}$ of terms that each contain only variables contained in a given maximal clique $C_l$.

One can then modify the optimisation problem in \eqref{eq:relaxapp} as follows:
replace the big $\Gamma^{(\nu)}$ matrix by a direct sum of smaller matrices $\Gamma_{l}^{(\nu)}$, one for each clique, constructed from the spin variables belonging to the clique $C_l$.
Some spin variables appear in more than one clique, which can be captured with additional linear constraints that involve variables of the blocks $\Gamma_{l}^{(\nu)}$.
Writing this explicitly, the chordal version of the SDP relaxation then reads as

\begin{align}\label{eq:relaxchordal}
 \min_{\Gamma^{(\nu)}_l} & \sum_n \tr( h_{n} \Gamma^{(\nu)}_{n} ) \nonumber \\
  \text{s.t.} {} \, & \,\,\, \Gamma_l^{(\nu)}  \succeq 0 , \quad \forall \, \, l = 1,\ldots n_C \, ,  \\
                         & \, \, \tr(F_{m,l} \Gamma_{l}^{(\nu)}) = 0 \, \, \, \, m = 1,{\ldots},k_l  \, \, ,  l = 1,\ldots n_C \, , \nonumber \\
                         & \, \,  \sum_{n} \tr((G_{l,n}) \Gamma_{n}^{(\nu)} ) = 0 \, \, \, \, l = 1,{\ldots},k  \, \, , \nonumber
\end{align}
where the $F_{m,l}$ are the intra-block constraints coming from the properties of the spin variables, while the $G_{l,n}$ correspond to the constraints identifying expectation values belonging to different blocks.
Interestingly this relaxation still converges to the exact result~\cite{lasserre2006convergent}.

Depending on the sparsity of the graph $\depgraphG$ (and its chordal extension $\depgraphG_C$), substituting the original optimisation relaxation \eqref{eq:relaxapp} by \eqref{eq:relaxchordal} leads to a substantial simplification and improved scaling in runtime and memory.
In practical applications, the latter are typically dominated by the the largest block, i.e., the largest maximal clique in $\depgraphG_C$.
Moreover, the block structure can be exploited to have a more finely tuned control of the lower bound precision, essentially by replacing a general hierarchy level $\nu$ by a moment matrix with block-dependent levels $\nu_l$. This allows to define hybrid levels where, for instance, smaller blocks are generated at higher values of $\nu_l$, thus improving the quality of the lower bound without significantly affecting the scalability of the SDP.

Let us illustrate how this chordal extended relaxation works in practice by going back to the three spins example introduced above and by considering the 1D Ising model with the Hamiltonian $H = \sum_{i =1}^2 J_{i,i+1} \sigma_i \sigma_{i+1}$ presented above.
The corresponding dependency graph $G$ is already chordal and is composed of two cliques $C_1 = {\sigma_1, \sigma_2}$ and $C_2 = {\sigma_2, \sigma_3}$.
Since the blocks at level $\nu = 1$ have been presented in the main text, here we show the relaxation at level $\nu = 2 $, where the matrix \eqref{eq:example} can be substituted by the two blocks:
\begin{align}
\Gamma_{C_1}^{(2)} &= \left( \label{eq:example_blocks1}
\begin{array}{cccc}
1 & \langle \sigma_1 \rangle & \langle \sigma_2 \rangle & \langle \sigma_1 \sigma_2 \rangle \\
 \langle \sigma_1 \rangle & 1 &  \langle \sigma_1 \sigma_2 \rangle & \langle \sigma_2 \rangle \\
\langle \sigma_2 \rangle & \langle \sigma_1 \sigma_2 \rangle &  1 & \langle \sigma_1 \rangle \\
 \langle \sigma_1 \sigma_2 \rangle & \langle \sigma_2  \rangle & \langle \sigma_1 \rangle  & 1 \\
\end{array}
\right) \\
\Gamma_{C_2}^{(2)} &= \left( \label{eq:example_blocks2}
\begin{array}{cccc}
1 & \langle \sigma_2 \rangle & \langle \sigma_3 \rangle & \langle \sigma_2 \sigma_3 \rangle \\
 \langle \sigma_2 \rangle & 1 &  \langle \sigma_2 \sigma_3 \rangle & \langle \sigma_3 \rangle \\
\langle \sigma_3 \rangle & \langle \sigma_2 \sigma_3 \rangle &  1 & \langle \sigma_2 \rangle \\
 \langle \sigma_2 \sigma_3 \rangle & \langle \sigma_3  \rangle & \langle \sigma_2 \rangle  & 1 \\
\end{array}
\right)
\end{align}
The constraints $G_{l,n}$ derive from the fact that the variable $\sigma_2$ belongs to both cliques, hence several expectation values appear in both blocks.
Some of the unnecessary expectation values in \eqref{eq:example}, such as $\langle \sigma_1 \sigma_3 \rangle$ and $\langle \sigma_1 \sigma_2 \sigma_3 \rangle$ no longer appear in the two smaller blocks $\Gamma_{C_1}^{(2)}$ and $\Gamma_{C_2}^{(2)}$.
Such a simplification is particularly useful because it reduces the number of variables involved in the SDP.

\section{Implementation of the chordal branch-and-bound method}
\label{app:CBBdetails}
We now describe in detail the implementation of our chordal branch-and-bound (CBB) method. First, we elaborate on the three ingredients: the lower bound, the upper bound, and the branching rule. Lastly, we comment on possible improvements of our strategy.

\subsection{Lower bound}

The cheapest method to get a bound on the ground-state energy from below is to use the relaxation in \eqref{eq:relaxchordal} at the lowest level $\nu$ that contains all the expectation values needed for the computation of the energy. In this work we focus on quadratic Hamiltonian functions, hence $\nu = 1$.
However, in practical applications we observe that this often leads to a lower bound that can be more than $10~\%$ away from the corresponding upper bound.
As already mentioned in Ref.~\cite{rendl2008solving}, having such a big initial gap slows down the convergence of the branch and bound, making it very difficult to reach a point in the branching where the lower bound is high enough to start excluding the first branches.
This problem is overcome by tightening the relaxation, which can be done in several ways.

The option considered in Ref.~\cite{rendl2008solving} was to tighten the relaxation at level $1$ by adding hand crafted linear inequalities, so-called triangle inequalities, between entries of the matrix corresponding to the two-body correlations $\langle \sigma_i  \sigma_j \rangle$ of triples of spin variables.
Since the amount of all these possible constraints scales as $N^3$, usually only a fraction of them is introduced.
In our CBB we can exploit the structure of the problem to obtain more systematic improvements.
The chordal extension reduces the amount of meaningful constraints that can be added.
Indeed, the resulting block structure implies that the only two-body expectation values $\langle \sigma_i  \sigma_j \rangle$ that appear in the moment matrices correspond to spins $i,j$ belonging to the same block.
Hence, all the triangle inequalities that can actually be imposed have to involve triples $i,j,k$ that appear in the same clique.

The numerical effort for one step in the CBB method is mostly determined by the largest block in the moment matrix.
Therefore, we choose to take a hybrid approach, introducing an intermediate level with $\nu = 2$ for all the blocks $\Gamma_l^{(\nu)}$ involving less than $n_t$ variables, while keeping all the bigger blocks at level $1$.
Taking such a hybrid level yields a significant improvement in the initial lower-upper bound gap already for smaller values of $n_t$. In fact, we have checked numerically that this devises a test that corresponds at least to the case of level $1$ plus the addition of all triangle inequalities between variables in the smaller blocks.

Moreover, we also allow for additional triangle constraints between two-body correlations belonging to bigger blocks.
In particular, we add them in an iterative way, as shown in Ref.~\cite{rendl2008solving}, until the improvement on the lower bound is smaller than some numerical precision.
In most cases we tested, there was actually no need to introduce these additional constraints.

\subsection{Upper bound}
For the upper bound one needs a good guess for a spin configurations that is close to the ground-state energy.
In our case, this is straightforward: from the moment matrices $\Gamma^{\ast (\nu)}_l$ obtained from the solution of the SDP \eqref{eq:relaxchordal}, construct the configuration $\vec{\sigma}^{\ast}$ where each spin $\sigma_i^{\ast}$ is aligned according to the sign of the entry corresponding to the expectation value $\langle \sigma_i \rangle$.
Intuitively, this can be seen as a way to obtain the spin configurations ``closest'' to the optimal (but typically unphysical) solution achieved by the relaxation.
A nice feature of our strategy for the upper bound is that it basically adds no extra computational cost as it is derived directly from the moment matrix that is obtained by solving the SDP.

The above method makes a substantial improvement over earlier approaches. Indeed, as described in Ref.~\cite{rendl2008solving},
the moment matrix used in previous SDP relaxations is missing the first row and column vector, and thus exactly the entries we need to extract our deterministic configuration.
That is why former approaches usually resort on performing a Cholesky decomposition of the moment matrices $\Gamma_l^{\ast} = B_l^{T} B_l$ and - by interpreting each row of the resulting matrices $B_l$ as a vector $v_{l,j}$ - assigning the deterministic values to the spin variables by taking scalar products of these vectors with a randomly chosen one.
Notice that, apart from requiring the additional computational effort of having to perform a Choleski decomposition, the above method is also hard to adapt to an SDP composed of more blocks, as the one resulting from the application of the chordal extension technique.

To conclude, once a valid spin configuration $\vec{\sigma}^\ast$ has been extracted, we simply set the upper bound to be its corresponding energy $H( \vec{\sigma}^\ast )$.
Surprisingly, we noticed that by following this procedure, the exact ground states is usually recovered very soon in the branching (see Fig.~1 in the main text for an example).
It then takes additional time to find a matching lower bound to verify that this is indeed the lowest achievable energy.
This makes us believe that our procedure is very efficient in finding the optimal deterministic configuration.

\subsection{Branching rule}
Here we follow the same method outlined in~\cite{rendl2008solving}, but with a different choice of branching procedure.
Indeed, the authors in~\cite{rendl2008solving} choose the dicotomic choice to be the relative alignment of a pair of connected spins.
That is, given a choice of indices $i,j$, the two branches correspond to the two cases $\sigma_i \pm \sigma_j = 0$.
However, as mentioned before, we prefer to branch on the value of the single spin, by choosing between the two values $\sigma_i  = \pm 1$.
The reason for this is that, in the latter case the number of possible branching steps depends only on the number $N$ of spins in the system.
On the contrary, the former method involves an amount of branching choices that depends on the number of edges in the dependency graph $\depgraphG$, which can be much higher, often as high as $N^2$.

For our branching rule strategy, there is the question of which spin $i$ to choose for the next branching.
The way we do this here is based on the expectation values $\langle \sigma_i \rangle$  recovered from the moment matrices $\Gamma_l^{(\nu)}$ and used for the construction of the upper bound.
The intuition is the following: spins with an expectation value close to zero are ``difficult'' choices, because flipping the value of such a spin is likely to lead to a slight change in the energy of the system; conversely, expectations values very close to $\pm 1$ are identified as ``easy'' choices, because flipping such a spin is likely to lead to a significant change in the energy of the system and it is easier to discard a branch during the evolution of the branch-and-bound process.
We set the branching rule to ``easy-first'', that is, at the end of each optimisation round, the next branching is performed on the closest to deterministic spin in the $\Gamma_l^{(\nu)}$.

\subsection{Possible improvements}

There is some freedom in the choices we outlined in the previous subsections.
Given the huge difference in complexity that can be exhibited by various instances of the Ising model, we expect the optimal choice to be model dependent.
Here we briefly discuss which modifications we imagine to be most useful for practical applications.

Let us start by recalling that, in order to accelerate the convergence process and to keep memory requirements low, it is crucial to reduce the initial lower-upper bound gap as much as possible and as early as possible.
One way to do that is to modify the hybrid hierarchy level introduced above.
In our applications, it was always enough to set the threshold to at most $n_t = 7$.
However, such value can be significantly increased without affecting too much the scalability of the CBB.
Indeed, the main bottleneck of our method is the memory required to solve the largest SDP.
This depends mainly on the block $l^{\ast}$ leading to the largest matrix $\Gamma_{l^{\ast}}^{(\nu)}$.
Therefore, as long as increasing the level of the smaller blocks does not lead to bigger matrices that the one for the largest clique, the SDP will still have the same memory requirements -- although the solving time will clearly increase.

Other branching rules can be also be adopted.
For instance, one can replace the ``easy-first'' approach with a ``difficult-first''.
In this case, one picks the next branching from the least deterministic spin in the $\Gamma_l^{(\nu)}$.
We expect the choice of the most effective branching rule to depend on the system under consideration.

Lastly, there are instances in which CBB does not outperform other methods. This is true for specific cases of very sparse problems, where linear-programming relaxations were shown to work very well~\cite{liers2004computing}, or some hand-crafted models for which exact polynomial algorithms are known~\cite{mandra2017pitfalls}.
Nevertheless, it would still be interesting to see if one could combine the construction based on the chordal extension with those methods and provide some further advantage.
\section{Phase transition in the ferromagnetic disordered Ising model in a 2D triangular lattice}\label{app:phasetrans}

\begin{figure}[t]
\centering
\includegraphics[width=\linewidth]{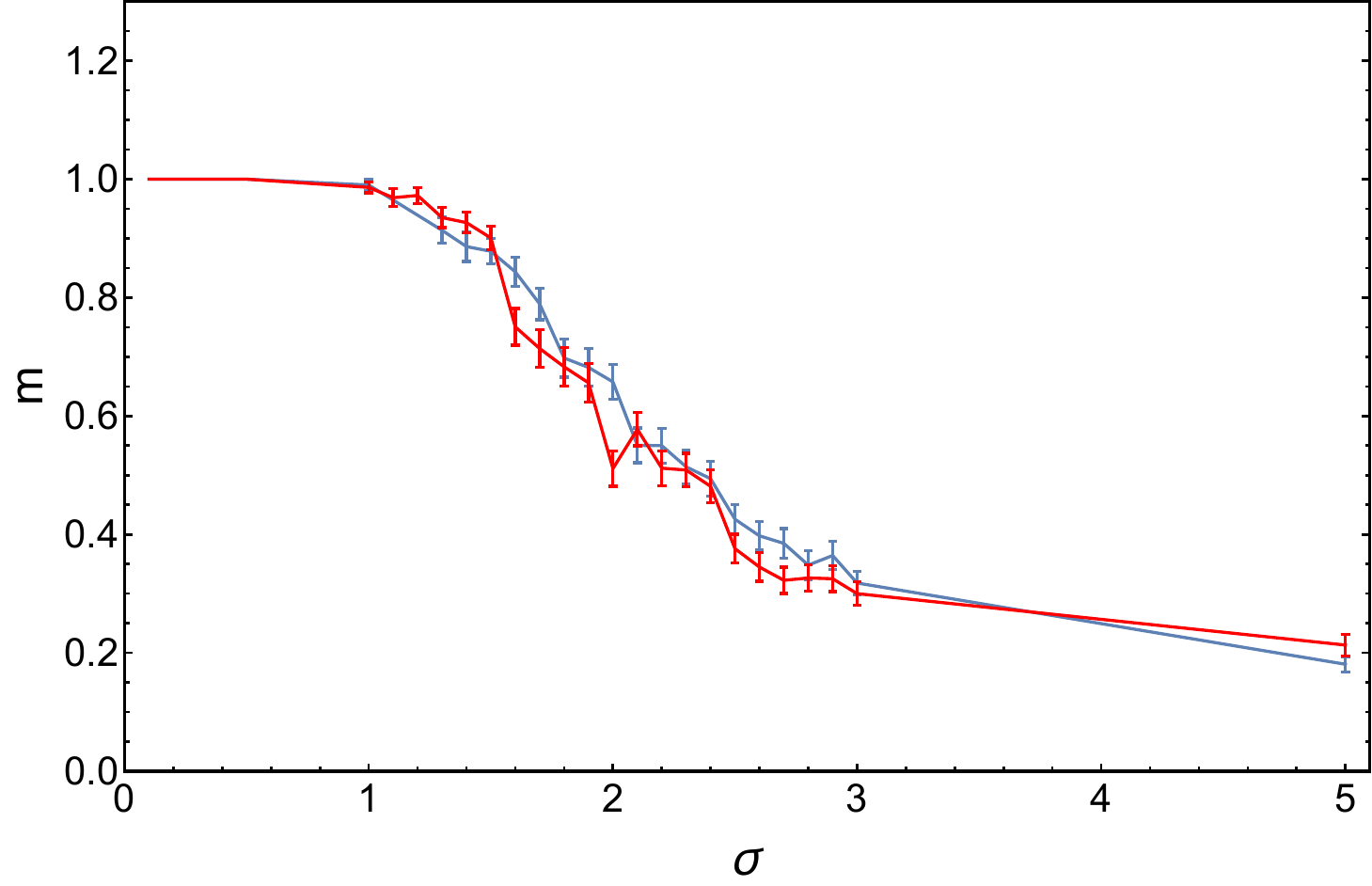}
\caption{Average magnetisation $m$ of the ground state of the ferromagnetic Ising model with disordered magnetic field, as a function of the disorder strength $\sigma$, for a 2D triangular lattice of linear size $L$. The plot shows the cases $L = 15$ (blue) and $L = 20$ (red).
  The phase transition could be pinned-down more precisely by means of a finite size scaling analysis.
  For our purposes here it is sufficient to know that in the range $\sigma \in [1,3]$, the ground states has non-trivial spin patterns.}
\label{fig:phasetrans}
\end{figure}

Here we use the CBB method to study in more detail the 2D triangular Ising model used to benchmark the solution of the D-wave quantum annealer. Recall that we considered a ferromagnetic model in a statically disordered mangetic field, represented by the following Hamiltonian 
\begin{equation}
H(\vec{\sigma}) = - \sum_{ \langle i,j \rangle } \sigma_i \sigma_j + \sum_{i = 1}^N h_i \sigma_i \, ,
\end{equation}
where the first sum runs over all pairs of connected spins $\langle i,j \rangle $ in the triangular lattice. The local fields $h_i$ are drawn randomly from a Gaussian distribution with zero mean and variance $\sigma$ for each site $i$.
As a function of the disorder strength $\sigma$, the model undergoes a phase transition from a ferromagnetic ground state (in which
all states are aligned with each other) to a disordered
phase (in which, for extremely large disorder, the spins
are aligned with the local magnetic fields).
The transition can be detected by estimating the ground state magnetisation
\begin{equation}
m = \frac{\mid \sum_{i = 1}^N \sigma_i^{(g)} \mid }{N} \, ,
\end{equation}
where $\vec{\sigma}^{(g)}$ is the ground state configuration.
Clearly, a ferromagnetic ground states is fully magnetized, hence it is characterized by $m = 1$. On the other hand, a disordered phase correpsonds to $m \approx 0$ (notice that, because of finite size effects, one can never reach an exactly zero value in numerical tests ).

We analyzed the phase transition by computing the ground state energy for the model with the CBB method and estimating its magnetisation $m$ as a function of the disorder strength $\sigma$. 
For each value of $\sigma$, the results were averaged over $100$ samples.
Figure \ref{fig:phasetrans} shows the obtained magnetisation curves, for two different linear lattice sizes $L = 15,20$.

\end{document}